\documentclass[%
 amsmath,amphysicalsymb,
 aps,
pra,
 reprint,
 nofootinbib,
 floatfix,
 longbibliography
]{revtex4-1}
\usepackage{graphicx}

\begin{document}

\title{Quantum-assured magnetic navigation achieves positioning accuracy better than a strategic-grade INS in airborne and ground-based field trials}

\author{Murat~Murado\u{g}lu}
\author{Mattias~T.~Johnsson}
\author{Nathanial~M.~Wilson} 
\author{Yuval~Cohen} 
\author{Dongki~Shin}
\author{Tomas~Navickas}
\author{Tadas~Pyragius}
\author{Divya~Thomas}
\author{Daniel~Thompson}
\author{Steven~I.~Moore}
\author{Md~Tanvir~Rahman}
\author{Adrian~Walker}
\author{Indranil~Dutta}
\author{Suraj~Bijjahalli}
\author{Jacob~Berlocher}
\author{Michael~R.~Hush}
\author{Russell~P.~Anderson}
\author{Stuart~S.~Szigeti}
\author{Michael~J.~Biercuk}

\affiliation{Q-CTRL, Sydney, NSW Australia}

\begin{abstract}

Modern navigation systems rely critically on GNSS, which in many cases is unavailable or unreliable (e.g. due to jamming or spoofing). For this reason there is great interest in augmenting backup navigation systems such as (drift-prone) inertial navigation systems (INS) with additional modalities that reduce positioning error in the absence of reliable GNSS. Magnetic-anomaly navigation (MagNav) is one such approach, providing passive, non-jammable navigation through periodic position fixes obtained by comparing local measurements of Earth's crustal field against known anomaly maps. Despite its potential, existing MagNav efforts have been limited by magnetometer performance and interference due to platform noise; solutions addressing these problems have proven either too brittle or impractical for realistic deployment. Here we demonstrate the performance of a quantum-assured MagNav solution based on proprietary quantum magnetometers augmented by a novel denoising and map-matching algorithm suite. The system is small enough to fit on fixed-wing drones or be integrated into the avionics bay of a commercial airliner. We present flight trials at altitudes from ground level to $\sim$19,000 feet, testing various configurations of onboard and outboard-mounted quantum magnetometers, and comparing against a strategic-grade INS. Our MagNav solution achieves superior performance, delivering up to $\sim$46$\times$ better (lower) positioning error than the velocity-aided INS; the best final positioning accuracy we achieve on a flight trial is 22m or $0.006\%$ of the flight distance. Airborne trials consistently achieve at least $11\times$ advantage over the INS across varying conditions, altitudes, and flight patterns. We demonstrate that the system can learn relevant model parameters online without special vehicle maneuvers, which provides robustness against various configuration changes (e.g. changing payload or latitude). Our trials also include the first successful MagNav performed in a ground vehicle using publicly-available anomaly maps, delivering bounded positioning error $\sim$7$\times$ lower than the INS, with both systems in strapdown configuration. 
 
\end{abstract}

\maketitle

\section*{Introduction}

Modern navigation systems depend heavily on GNSS due to several attractive features that benefit a wide variety of missions~\cite{Bernhard2008}. First, it is passive with regard to the receiver, allowing the user to locate themselves without revealing their location~\cite{Montenbruck2017}. Second, the beacons that are used, satellites, are readily accessible across the globe, enabling a single navigational system to be used in almost any location~\cite{Montenbruck2017}. Despite its ubiquity, GNSS has increasingly become an area of strategic concern, due to poor coverage in some strategically-relevant regions (such as near the poles) or in critical operational domains (such as underwater). In addition, GNSS jamming and spoofing has become a major geopolitical concern due to disruptions to both global trade and military operations~\cite{royal2011global}. 

Alternative navigation solutions have struggled to compete with the exceptional advantages of GNSS, despite its emerging vulnerabilities. The standard approach to navigation with unreliable GNSS signals is to rely on inertial navigation systems (INS), which integrate accelerations and rotation rates over time to calculate a navigation solution. Such systems are passive and operational anywhere, but suffer from position errors that grow over time without bound~\cite{titterton}. Higher grades of INS can slow the error growth rate, but cannot eliminate it. This means that for long-duration navigation~\cite{titterton}, the INS estimate must be augmented with some additional navigation aid. 

A multitude of beacon-based aiding technologies have been developed that leverage existing infrastructure (e.g. mobile phone towers) and/or involve the installation of an extensive array of custom beacons across an operating region~\cite{Son:2018, zafari2019surveyindoorlocalizationsystems}. However, these can also be subject to jamming and spoofing. In contrast, beacon-free active navigation solutions relying on Doppler radar~\cite{RadarDawson2022} and lidar~\cite{LIDARZhang2017} provide greater positioning resilience, albeit at the significant expense of potentially revealing a user's location.  

Developing a GNSS alternative that is both passive and that does not require active beacons is a pressing challenge, with various partial solutions under investigation;  many such systems rely on various forms of optical technology. For instance, camera-based visual terrain recognition~\cite{Bijjahalli:2020} and star cameras~\cite{groves2013} are now available commercially, but have common vulnerabilities that limit their operation: these approaches fail under poor visual conditions, caused by weather or the time of day, or when operated in featureless areas such as oceans and deserts~\cite{Balamurugan2016SurveyOU}.

Magnetic-anomaly navigation (MagNav)~\cite{canciani2016} provides a beacon-free position fix which is not subject to the same limitations; it is available in all geographies and is not affected by weather, visual obstructions, or time of day. It is also completely passive and largely unjammable/unspoofable, giving it significant advantages for operation in contested environments. 

MagNav relies on the fact that the Earth’s magnetic field possesses small amounts of local variation (``anomalies'') that are geographically distinct.  These anomalies are stable in time and have been mapped for various purposes, including resource exploration. With an appropriate sensor capable of detecting these anomalies, and an available reference map, it is possible to infer positional information to both improve on INS positioning and provide bounded accuracy indefinitely. 

The magnetic field that is measured is composed of several parts. There is the core Earth field, which is described by a time-dependent model such as the International Geomagnetic Reference Field (IGRF) \cite{IGRF}, and has a scalar magnitude of $\sim$25,000 -- 65,000 nanotesla (nT). On top of this there are anomalies that arise from crustal geology and are stable in time. These variations are on the order of 10\,nT -- 100\,nT over a few kilometers and are what is used for MagNav. Global anomaly maps have been produced, such as the Earth Magnetic Anomaly Grid Version 3 \cite{EMAG2v3} or the World Digital Magnetic Anomaly Map (WDMAM) \cite{WDMAM}, and can in principle be used for navigation. These are well-supplemented by higher-resolution maps developed by the geophysical surveying sector or defense agencies. Finally there are time-dependent effects such diurnal ionospheric variation ($\sim$100\,nT) and space weather arising predominantly from solar activity (up to 1000s of nT during solar events) \cite{langel1998}.

Providing a position fix via MagNav for GNSS-free navigation has been studied extensively over decades. Most of this work consists of results purely in simulation, for example Refs.~\cite{mcneil2022,greentree2023,gupta, tkhorenko}, although some real-world flight-test results exist~\cite{wilson2006,cancianiMapQuality, canciani2, canciani-f16, leecanciani}. Even with this work, MagNav has until now faced major challenges in delivering mission-relevant capabilities. These include:
\begin{itemize}
    \item Development of small-form-factor magnetometers with sufficient sensitivity and stability to enable anomaly identification without suffering from drift, noise, significant heading errors, or data instabilities;
    \item Mitigation of platform noise, where magnetic interference can be $100-1000\times$ larger than target magnetic anomaly magnitudes, reducing MagNav accuracy;
    \item Development of sensor-fusion algorithms to extract optimum information from noisy magnetometer measurements and fusing this data with other onboard navigation systems such as INS position estimates;
    \item Reducing the need for extensive calibration of algorithms in the form of requisite vehicle maneuvers (e.g. special cloverleaf or box maneuvers at the start of every flight) or extensive training-data-acquisition prior to commencing MagNav;
    \item Combating the intrinsic brittleness of positioning algorithms against vehicle, payload, latitude, altitude, and sensor-location changes; and
    \item Accounting for imperfections arising from adjusting anomaly maps for altitude and integrating realistic time-varying effects such as space weather and diurnal magnetic field variations.
\end{itemize}

In this paper we address these key challenges through the development of a novel approach to quantum-assured MagNav, combining hardware and software-level innovations. Ultimately, these innovations are demonstrated through both airborne and ground-based field trials that validate the quality of solutions in real operating environments of interest to end users.

\begin{figure*}
\center
\includegraphics[width=0.9\textwidth]{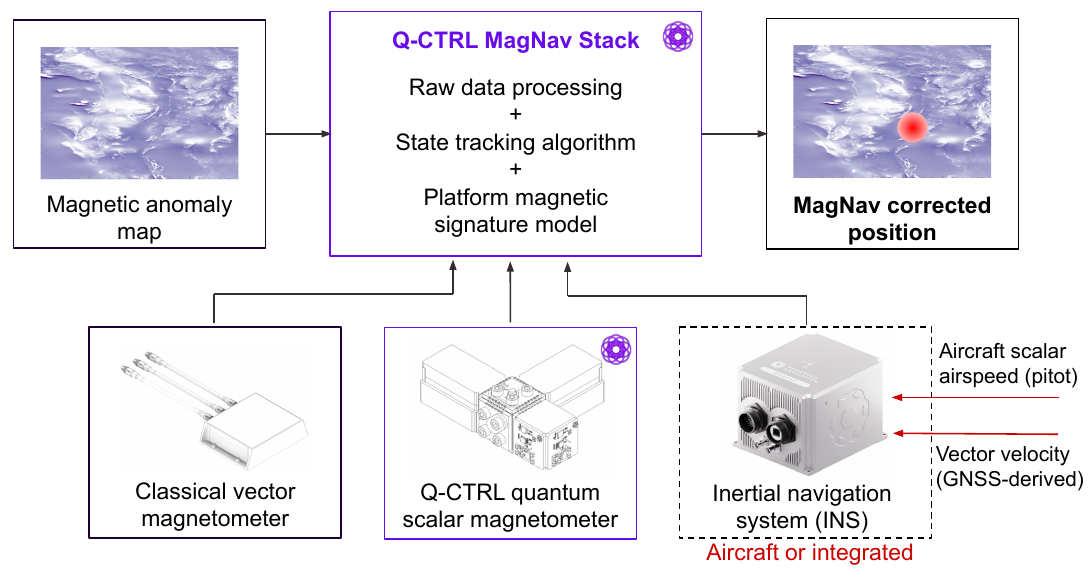}
\caption{System architecture of the MagNav system used in the trials. Boxes with purple borders indicate Q-CTRL produced components. 
 Black boundary boxes indicate third-party inclusions (e.g. the `classical' vector fluxgate magnetometer). We include an INS in our test system, but this may be optionally replaced with a feed direct from the vehicle's flight-management system. In the flight tests presented below we leverage different airspeed-sensor inputs as a part of system validation; ultimate GNSS-denied performance would only incorporate signals from onboard airspeed sensors. The  INS used in the ground trials had no source of velocity aiding.}\label{fig:system_architecture}
\end{figure*}

\section*{Q-CTRL MagNav System}
\subsection*{System architecture}
Q-CTRL has developed a full-stack solution for MagNav combining in-house-developed hardware and various software-level innovations (Fig.~\ref{fig:system_architecture}). We have designed a novel architecture that delivers maximum performance in a small form factor; key subsystems include a scalar quantum magnetometer, a vector classical magnetometer, a classical INS, and a magnetic denoising and map-matching algorithm suite. The system takes as inputs a magnetic map, vehicle-based-INS signal (optionally integrated into the MagNav system, as tested), and vehicle-velocity information. Sensor inputs are processed in combination with these data streams in order to provide an iteratively-updated estimate of the MagNav-corrected position.

We employ commercial-off-the-shelf (COTS) sensors for the vector magnetometer (Bartington Mag619U) and for the INS. The INS is a high-end strategic-grade system with IMU specifications given in Appendix Table~\ref{tab:INSspecs}. In the subsections below, we provide further details on the Q-CTRL subsystems (see items with purple boundaries in Fig.~\ref{fig:system_architecture}).

\subsection*{Q-CTRL optically-pumped magnetometers}
\begin{figure*}[ht]
\centering
\includegraphics[width=1.0\linewidth]{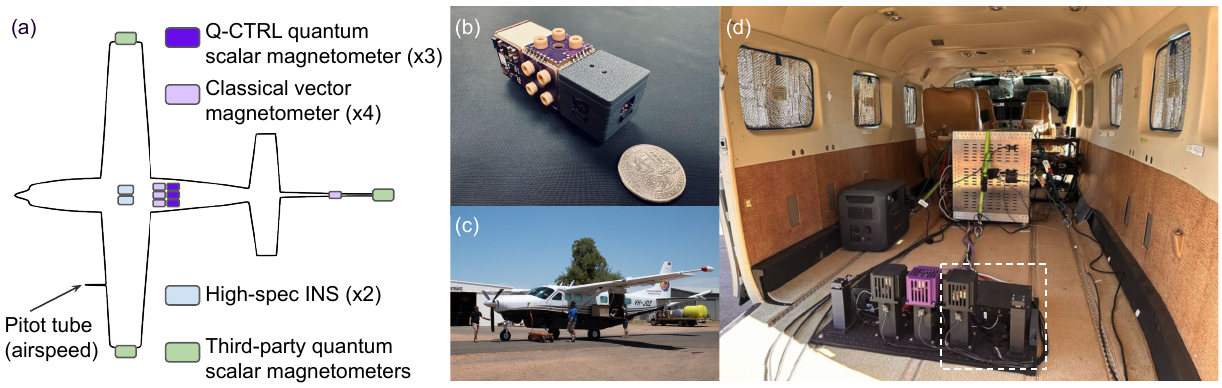}
\caption{(a) Schematic showing the instrumentation layout on the flight trial aircraft. The Q-CTRL quantum magnetometers were mounted internally in a high noise location. Additional magnetometers were placed externally in order to obtain ground truth and for comparison purposes. The pitot tube provided a scalar velocity input for INS aiding. (b) Photograph of a single colinear Q-CTRL scalar quantum magnetometer, shown alongside a US quarter-dollar coin for a sense of scale. (c) Cessna 208B Grand Caravan used for the trials. (d) Positioning of the Q-CTRL sensor package, power supply, control electronics and logging equipment inside the aircraft. This is a triple-redundant configuration employed for testing and validation; a single scalar/vector magnetometer pair comprises a complete MagNav hardware system. The dashed box shows a potential minimal configuration consisting of one Q-CTRL scalar magnetometer, one fluxgate vector magnetometer, and the ancillary control electronics, with a total volume of 4.2\,L. The same onboard system operated in strapdown configuration is employed in ground-based trials.}
\label{fig:aircraft_schematic_three_column}
\end{figure*}

The Q-CTRL magnetometers are scalar optically-pumped magnetometers based on optical detection of atomic spin precession using a vapor cell containing rubidium atoms in a buffer gas~\cite{Budker:2013}. All photodetectors and light sources are integrated into the sensor head, along with vapor cell heaters (Fig.~\ref{fig:aircraft_schematic_three_column}b). We have produced variants with pump and probe beams in both orthogonal and co-linear geometries;  performance is comparable other than the modest increase in size associated with the orthogonal optical configuration. For the field trials conducted here, we employ the orthogonal configuration. The sensor head is connected to a single integrated electronic controller board via a ribbon cable, with power supplied externally.  

The Q-CTRL magnetometer sensor head achieves an absolute sensitivity of $<$80\,fT/$\sqrt{\mathrm{Hz}}$ at Earth's field, with a bandwidth of 250\,Hz. Each weighs approximately 70\,g, has a volume of 144\,cm$^{3}$, and consumes less than $15$\,W of power (limited by current electronics).  These magnetometers have undergone laboratory vibration tests to $5.7g$, and have been validated to provide a stable output in the operational range of $[-30, 50]^{\circ}$C.

\subsection*{Q-CTRL magnetic denoising and map-matching algorithms}
Q-CTRL has developed a comprehensive software stack incorporating a map engine and a navigation-and-map-matching engine, as shown in Fig.~\ref{fig:system_architecture}. The map engine includes core and anomaly field modelling, map levelling, upward and downward continuation, and prediction of temporal effects such as the diurnal variation and space weather. The navigation-and-map-matching engine includes platform denoising, statistical filters, and navigation algorithms. Our unique approach combines the processes of denoising and map-matching into a single step, rather than segregating those processes.

Core to this stack is the iterative filter used to calculate navigational corrections based on the (implicit) denoised magnetometer signals. The filter is designed to operate in real-time, enabling corrections to be provided in flight when connected to a flight-management system (i.e. it does not involve fitting or optimization over complete magnetometer data sets). In operation, filter updates can be processed up to 250\,Hz using standard embedded microprocessors.

Our approach to magnetic denoising is augmented by a physics-driven model used to learn the platform’s magnetic behavior and how it corrupts the external Earth field. This is performed by solving (and adaptively updating in real-time) a set of coefficients that provide a model of the vehicle’s magnetic field. These coefficients can act similarly to the Tolles-Lawson coefficients \cite{tolles1950, gnadt2022}, but are not strictly the same; their values initially fluctuate as the filter refines its estimate of the platform’s characteristics, but then largely settle and drift slowly over the navigational mission. Throughout the process, the vector and scalar magnetometer data is combined in conjunction with the current estimate of the platform characteristics in order to both remove platform noise and improve positioning accuracy.  

The algorithm initially has no knowledge of the detailed magnetic characteristics of the vehicle, other than plausible physical assumptions that are true for any vehicle. It rapidly learns the platform characteristics as the navigational mission begins, and this training is continuously refined. These real-time adaptive updates do not require any specific banks, turns, or deviations from the planned flight path. That is, rather than a pure data-driven approach requiring training against specially-designed calibration maneuvers beforehand, the algorithm we present is able to begin with zero training and adaptively updates its estimate of the platform’s magnetic characteristics throughout the mission. The platform model can be learned online in a specific mission as above (``cold start'') or recycled from a prior mission on the same vehicle for accelerated learning and immediate activation of MagNav (``warm start'').

This choice of filter architecture gives several significant benefits beyond cold-start operation. Most importantly, it allows the algorithm to adapt to significant changes in the platform’s configuration over time, imparting robustness to changes in payload, flight path, and latitude. The magnetic navigation algorithm is initialized with a starting position that need not include GNSS-level precision; tests show that the algorithm can tolerate initial position errors up to 4 km. Another strength of our physics-based approach is that it works ``out of the box'' at different latitudes or with different vehicle payloads without retraining. The iterative update process implicit to the algorithm continuously refines the model of the platform characteristics, enabling it to capture changing model parameters associated with e.g. changing latitude;  this contrasts significantly with data-driven approaches whose learned parameters are locked to a specific calibration conducted under a specific set of platform and environmental conditions.  

Position fixing is conducted via comparison to reference maps consumed by the MagNav algorithm. It leverages the IGRF-13 model \cite{IGRF} for the core field and EMAG2v3 model \cite{EMAG2v3} for the crustal anomaly map reference. However, the algorithm has the ability to add other higher-resolution local anomaly maps as layers; the navigation stack will use the most accurate selection in any given location. The resultant map field (core + anomaly) is then upward or downward continued to the altitude of the platform. 

\begin{figure*}[htp]
\begin{minipage}[b]{\linewidth}
\centering
\includegraphics[width=0.8\linewidth]{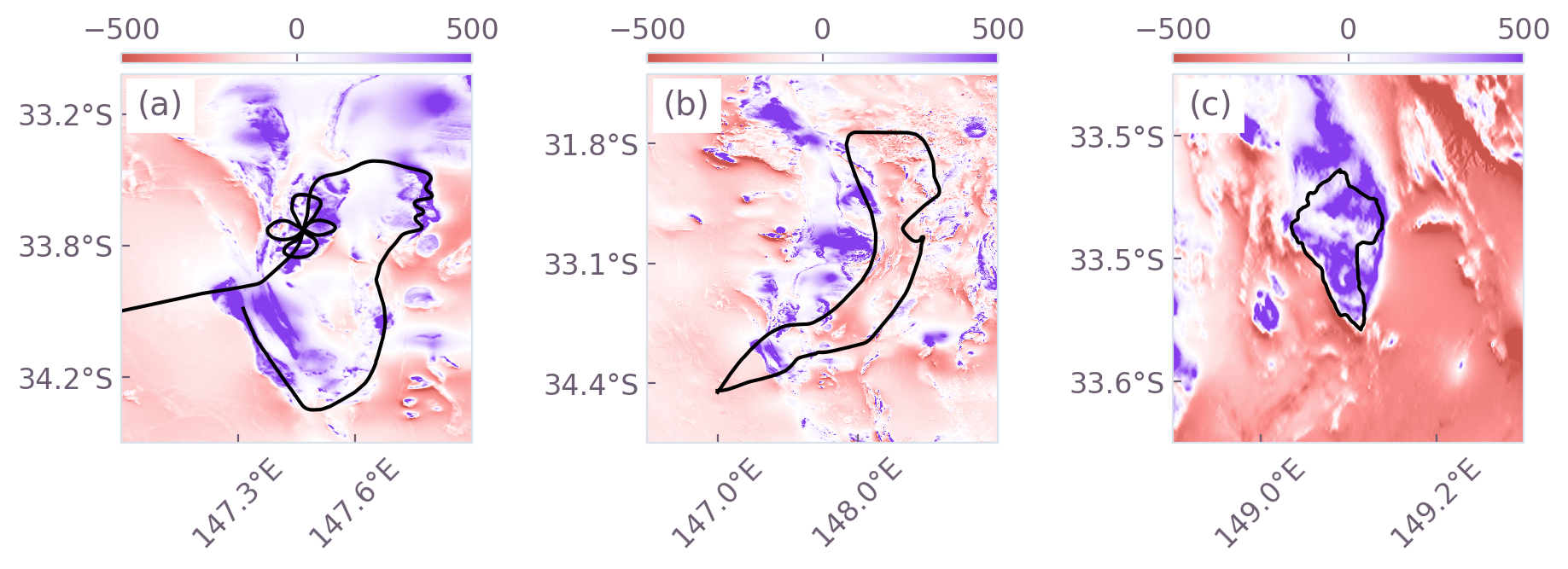}
\end{minipage}
\caption{Ground-truth trajectories superimposed on a magnetic anomaly map for (a) flight trials at 3600 feet, (b) flight trials at 19,000 feet and (c) ground-based trials. All magnetic anomalies are measured in nT.}
\label{fig:trajectory_maps_all}
\end{figure*}

\section*{Field trials}
Q-CTRL has performed a series of field trials in which its MagNav hardware and software were tested in real operating conditions in airborne and ground-based settings.  These trials involved exploration of environmental and operational tolerances, validation of hardware performance, and demonstrations of positioning capability.  Field-trial locations were selected for operational convenience near airfields and in areas with low vehicular traffic.  We did not select for locations with unusually strong magnetic anomalies, and flight paths traverse both regions with weak and strong magnetic anomaly relief.  

All field trials are performed using a ``demonstrator'' MagNav system which includes triple redundancy in the magnetometers and GPS receivers used to establish the ``ground truth'' in trials, seen in Fig~\ref{fig:aircraft_schematic_three_column}.  In addition, for certain tests we leverage outboard Geometrics optically-pumped magnetometers as an independent validation of our software and as a means to test functionality in low-magnetic-noise environments. 

During all trials we record signals from all magnetometers, the GNSS receivers, and the INS in order to calculate position estimates and then errors relative to ground-truth. The GNSS signal is used purely to obtain the ground-truth reference against which we can compare our navigation solution, and is not used for actual navigation. When performing MagNav, a single scalar magnetometer is used in conjunction with one vector magnetometer, and no cross-correlation or averaging is performed over the redundant hardware.

In all trials we compare MagNav performance against an INS as the most relevant competitive classical technology. We select this classical comparison for several reasons. First, it is the only classical alternative that operates under the same conditions as MagNav: compatible with flight platforms, and is operable independent of weather and visibility. Second, inertial navigation possesses similar advantages to MagNav: completely passive and undetectable, unjammable, and unspoofable. Finally, the strategic grade INS we employ has a similar size, weight and power as the complete MagNav solution.

\subsection*{Flight trials}
Over a week in February 2025, Q-CTRL conducted flight trials near Griffith, 
Australia. More than 6700\,km kilometers were flown using a Cessna 208B Grand Caravan at altitudes that ranged from near ground level up to 19,000 feet. Flights were conducted in various patterns. We note explicitly that the choice of certain trajectories, such as a cloverleaf pattern, is made to stress the INS and MagNav solution. We do not ever execute the formal Tolles-Lawson calibration patterns~\cite{gnadt2022} (flight along the sides of a box, with each side having chained roll / pitch / yaw maneuvers, followed by a cloverleaf for verification). Atmospheric conditions ranged from clear to the presence of electrical storms that required flight diversion.

The aircraft was equipped with a full suite of magnetometers in a variety of locations, as shown schematically in Fig.~\ref{fig:aircraft_schematic_three_column}. The actual Q-CTRL sensor package and ancillary equipment is shown in its operating configuration in the aircraft cargo bay. In these tests, we took advantage of the additional sensor inputs for test and validation purposes. For example, the presence of outboard stinger-mounted quantum optically-pumped magnetometers provided a ``low-noise'' reference against which the Q-CTRL onboard sensors could be compared.  

We use the system-integrated INS, though in principle a direct data feed from the aircraft's INS may be exploited. The INS was not provided with GNSS input, but it was provided with velocity-aiding to enable its peak performance. 

The results we present here focus on the most demanding test scenarios, where we primarily rely on the use of magnetometers mounted internally in the aircraft and engage the aircraft autopilot in order to evaluate the MagNav system performance in the highest-noise environment. We also include the vector-velocity-aided INS data to provide the ``best case'' INS performance comparison. Nevertheless, our trial data include a wide variety of comparisons tabulated below.

\subsection*{Ground trials}
Over three days in November 2024, Q-CTRL conducted land trials 30 minutes south of Orange, NSW, Australia, starting at Forest Reefs. 
In ground-based field trials the Q-CTRL system is loaded in strapdown configuration in the cargo bay of a standard rental van; an INS was provided as part of the test solution. This vehicle is not chosen to possess any specific low magnetic noise characteristics, and does not include any vibration compensation systems, making this road trial a test in extreme environments exhibiting the worst anticipated vibration and magnetic behavior. Typical routes were between $15-20$\,km in duration and traversed areas with varying magnetic anomaly relief with a mixture of sealed (75\%) and unsealed gravel roads (25\%) through winding hills. The elevation of this area was between $750-950$\,m.

\section*{Demonstration of MagNav}
In all cases we establish the ground-truth reference position using GNSS, and perform MagNav against public domain magnetic anomaly maps drawn from Geoscience Australia's ``Total Magnetic Intensity (TMI) Grid of Australia 2019 - 7th Edition'' \cite{TMIAustralia}. Our core test involves calculating positioning estimates via the Q-CTRL MagNav stack, and comparing performance against the positioning accuracy achieved using the INS co-located with the MagNav system.  Relevant magnetic maps overlaid with the ground-truth trajectory information are presented for several representative trials in Fig.~\ref{fig:trajectory_maps_all}.

\begin{figure}
\center
\includegraphics[width=\columnwidth]{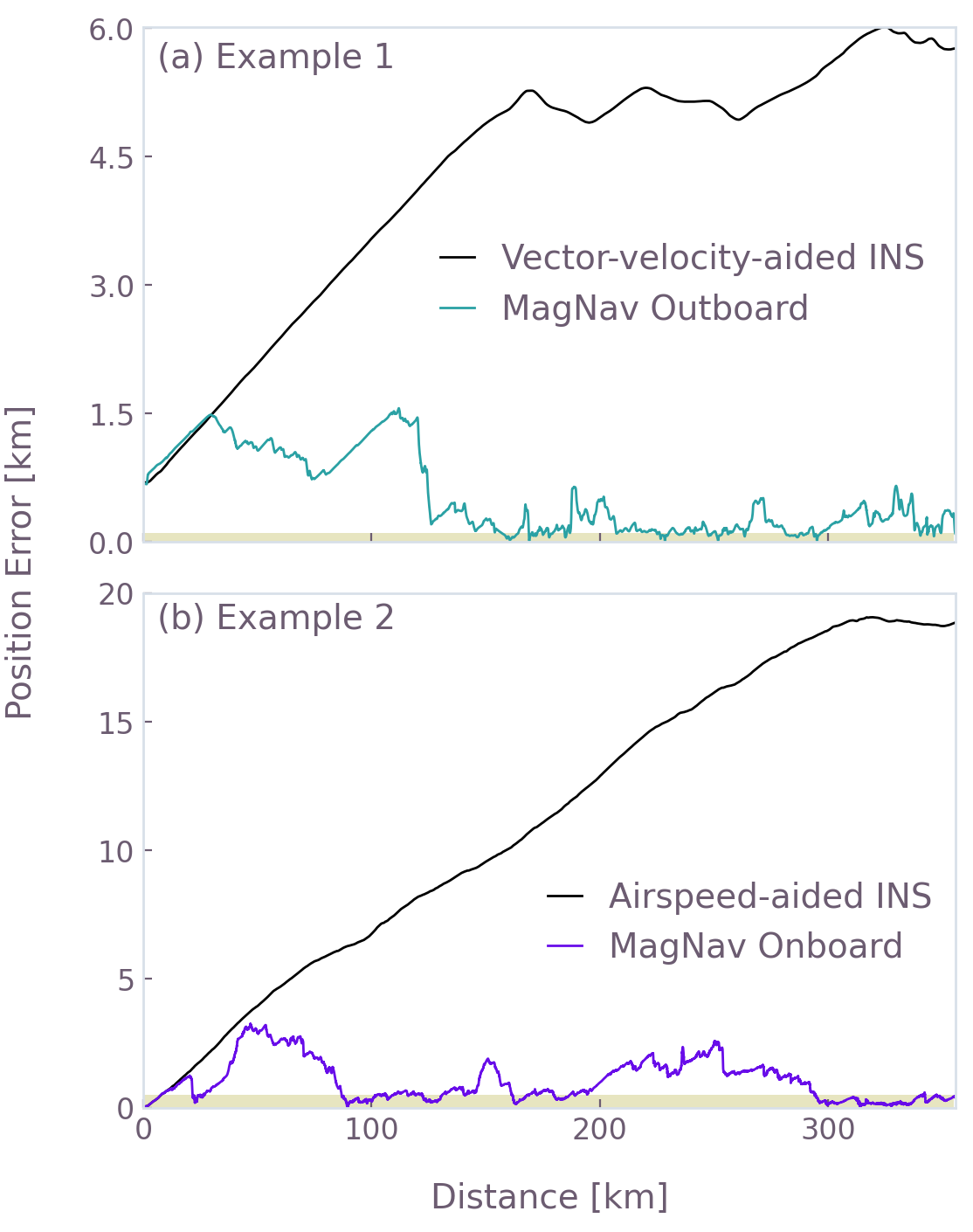}
\caption{Demonstration of quantum MagNav achieving a better bounded positioning accuracy when compared against different INS configurations for different flights at 3600 feet altitude, following the approximate trajectory illustrated in Fig.~\ref{fig:trajectory_maps_all}a. Note that these data are examples only and do not constitute a head-to-head comparison under identical conditions. (a) Comparison of positioning error using a vector-velocity-aided INS and the Q-CTRL MagNav system. Outboard externally-mounted quantum magnetometers in use. Shaded horizontal band represents 100m accuracy. Relative advantage over the INS at the conclusion of the flight is $\sim$46$\times$. (b) Similar comparison using onboard quantum magnetometer and comparison against INS aided by scalar airspeed. Shaded band represents 500m accuracy. Relative advantage over the INS at the conclusion of the flight is $\sim$34$\times$.}
\label{fig:3dvelocityaid_outboardmag_3600ft}
\end{figure}

\begin{figure*}[htb]
\begin{minipage}[b]{\linewidth}
\centering
\includegraphics[width=0.7\linewidth]{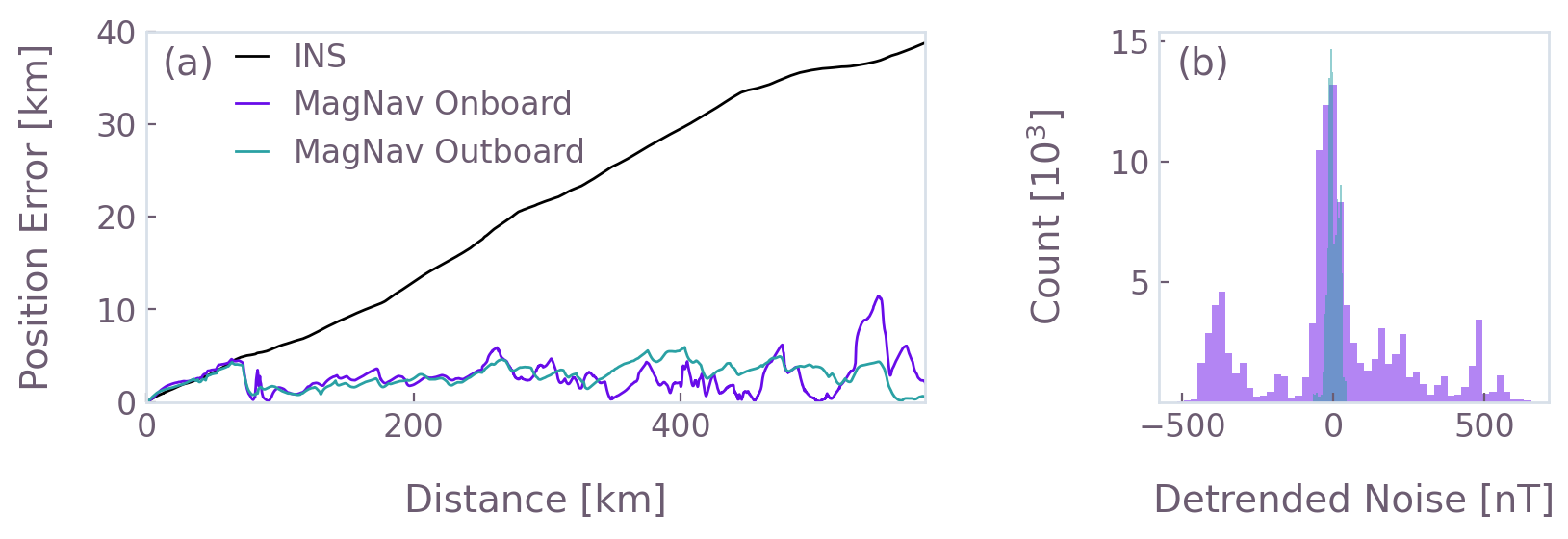}
\end{minipage}
\caption{Comparison of MagNav performance using onboard vs outboard (stinger-mounted) quantum scalar magnetometers in flight. Flights were conducted at 19,000 feet with the autopilot engaged to increase the magnetic noise, following the trajectory illustrated in Fig.~\ref{fig:trajectory_maps_all}b. (a) Positioning-error for both sensor configurations as compared against a scalar-airspeed-aided INS. (b) Histogram of the magnetic noise over the flight, comparing the noise environment of the onboard magnetometer (purple) and the stinger-mounted magnetometer (cyan). The RMS magnetic noise recorded by the onboard magnetometer is approximately $13\times$ larger than that mounted in the stinger. }
\label{fig:position_error_onboard_vs_outboard_19000ft}
\end{figure*}

\begin{figure}
\center
\includegraphics[width=\columnwidth]{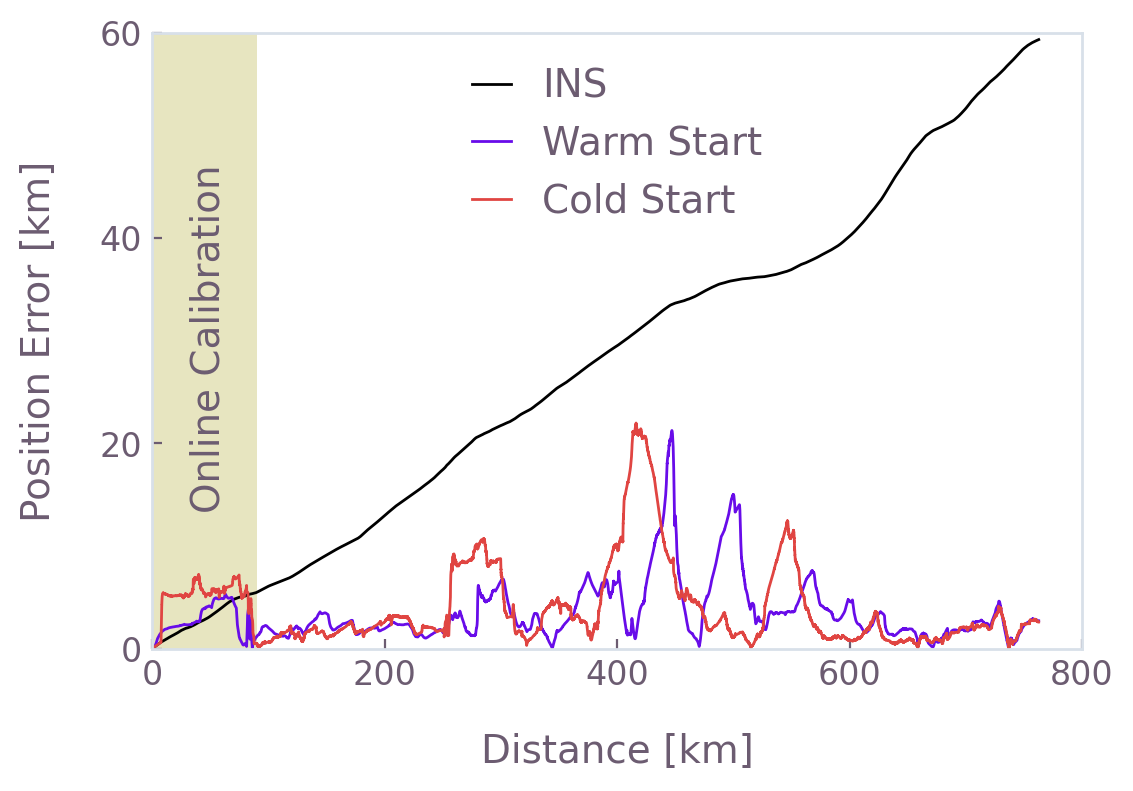}
\caption{Magnav results for a high-altitude flight at 19,000 feet with a total flight distance of 750\,km, comparing warm-start and cold-start modes. In the warm-start mode, the system was initialized using results from a previous flight, allowing pre-calibration of the magnetic signature model of the aircraft. In the cold-start mode, the system began with no knowledge of the platform characteristics. No calibration maneuvers were performed. Magnetic anomaly navigation performance is statistically indistinguishable over the flight.}\label{fig:cold_vs_warm}
\end{figure}

Flight trial results are presented across Figs.~\ref{fig:3dvelocityaid_outboardmag_3600ft}--\ref{fig:cold_vs_warm}, demonstrating that the quantum-assured MagNav solution can outperform the INS across a wide range of conditions. Figure~\ref{fig:3dvelocityaid_outboardmag_3600ft} presents two key examples which validate the advantage of MagNav, and show that throughout the flight the MagNav solution provides bounded positioning accuracy that does not grow with flight duration. Note the two panels do not constitute a direct comparison, despite a similar appearance. First, Fig.~\ref{fig:3dvelocityaid_outboardmag_3600ft}a demonstrates that the largest advantage we achieve ($\sim$46$\times$) is realized using externally-mounted quantum magnetometers, referenced against a vector-velocity-aided INS. In this flight, conducted at 3600 feet altitude, the ultimate positioning error is just $\sim$112\,m, or approximately $0.024\%$.  Figure~\ref{fig:3dvelocityaid_outboardmag_3600ft}b demonstrates that significant advantage for MagNav is also achievable under very different conditions: complete GNSS denial, reference against an airspeed-aided INS, and use of the onboard Q-CTRL magnetometer. In both of these examples the MagNav positioning accuracy quasi-periodically drops below the $\sim100$~m scale, indicating that throughout a flight, tightly bounded positioning is possible.

The best final positioning accuracy we achieve using MagNav is 22m or $0.006\%$ of the flight distance, outperforming a vector-velocity-aided INS by $\sim15\times$ (see Table~\ref{tab:results}). This result is nearly an order of magnitude better than the typical results achieved using active GNSS alternatives such as Doppler radar or Doppler velocity lidar. Typical positioning uncertainty in complete GNSS denial achieved at the end of a $\sim$365~km flight ranges from $\sim$ $0.14-0.6\%$, and similar performance is achieved at altitudes up to 19,000 feet.  

\begin{table*}[h!t!b]
    \centering
    \begin{tabular}{|c|c|c|c|c|c|c|c|c|c|c|}
\hline
Trajectory  & Altitude      & Scalar    & Velocity    & Distance  & \multicolumn{2}{c|}{INS Error}    & \multicolumn{2}{c|}{MagNav Error} & Advantage  \\
            & (ft)          & Mag.      & Aid         & (km)     & (m)          & \%                 & (m)        &  \%                  & Factor \\
\hline
(a)         & 3,600	      & onboard	  & airspeed    &365        & 19,313	   & 5.3\%	            & 515	     & 0.14\%	            & 38     \\ 
(a)*         & 3,600	      & onboard  & airspeed    &365        & 19,313	   & 5.3\%	            & 566	     & 0.16\%	            & 34     \\ 
(a)         & 3,600	      & onboard	  & 3D velocity &365        & 319          & 0.09\%	            & 22	     & 0.006\%	            & 15     \\ 
(a)$^\dag$  & 3,600	      & outboard  & 3D velocity &420	 	& 5,125	       & 1.2\%              & 112        & 0.027\%			    & 46     \\ 
(b)	      & 16,000–18,000 & outboard  & airspeed    &580        & 40,000       & 6.9\%	            & 2,000	     & 0.34\%			    & 20     \\ 
(b)	      & 15,000–19,000 & onboard	  & airspeed    &300	    & 40,000	   & 13.3\%		        & 2,000	     & 0.67\%			    & 20     \\ 
(c)         & ground        & onboard	  & NA          &18         & 1200         & 6.9\%	            & 180	      & 1.03\%	            & 7      \\ 
\hline
    \end{tabular}
    \caption{Representative trial performance characteristics and positioning accuracy. Trajectories are shown in Fig.~\ref{fig:trajectory_maps_all}.*: Flight trial compares cold-start filter to warm-start. $^\dag$:Trajectory was largely the same as (a) but had some minor deviations due to weather conditions.}
    \label{tab:results}
\end{table*} 

We achieve $11$-$38\times$ better position accuracy relative to the INS using internally-mounted quantum magnetometers across different flights, demonstrating that even in an environment with roughly an order of magnitude higher noise inside the aircraft, the Q-CTRL denoising and map-matching algorithms are able to faithfully calculate accurate navigation corrections. We can directly evaluate the efficacy of our denoising engine by comparing achievable positioning accuracy using onboard (high-noise environment) and outboard (low-noise environment) scalar quantum magnetometer signals with all other conditions held fixed. A demonstration of such results is shown in Fig.~\ref{fig:position_error_onboard_vs_outboard_19000ft}, together with the magnetic noise environment and the flight map. The internally-mounted magnetometer signal is approximately $13\times$ noisier than the outboard stinger-mounted signal, but we achieve comparable positioning performance. We note that the absolute performance achieved in this trial may be bounded by the anomaly features and small differences in the outboard versus onboard quantum magnetometers, but the ability of our navigational software stack to efficiently suppress noise is evident.

A summary of the key performance metrics for each flight field trial is provided in Table~\ref{tab:results}. Across all trials, the quantum-assured MagNav solution outperforms the INS.

\begin{figure}
\center
\includegraphics[width=\columnwidth]{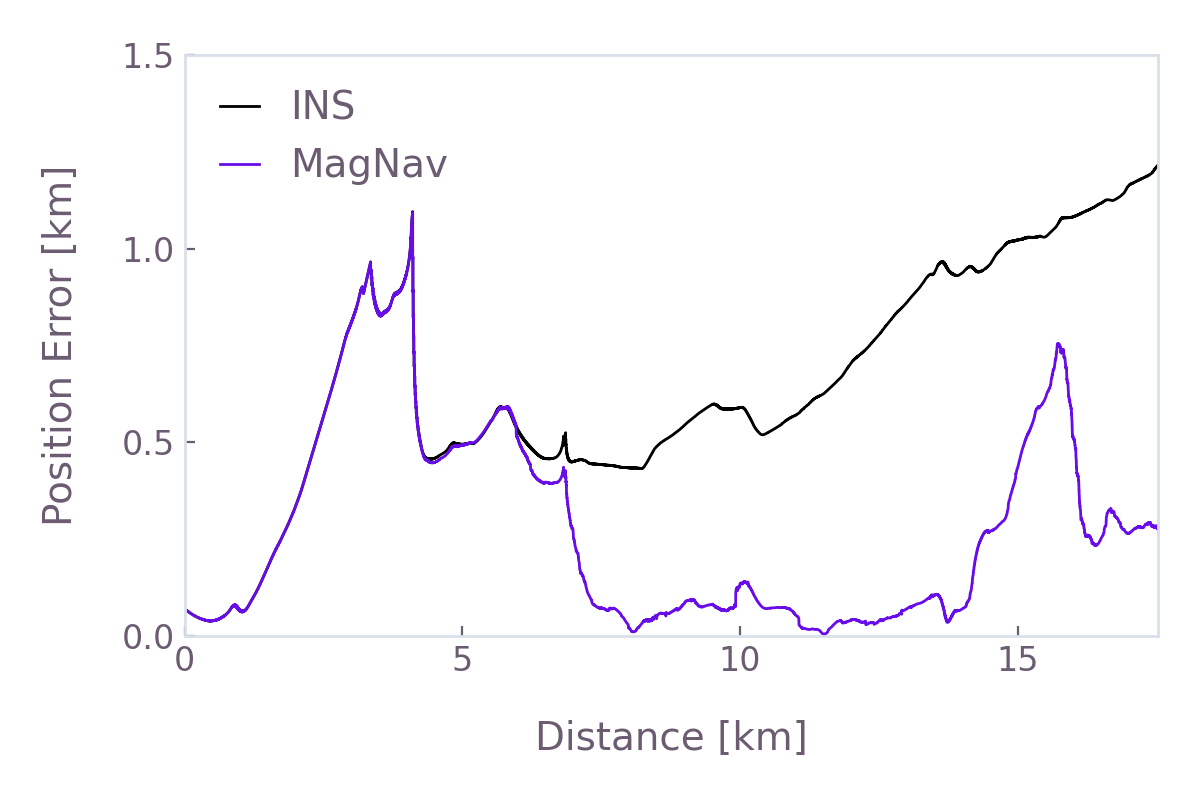}
\caption{Ground-based trial data.  Positioning error of the Q-CTRL MagNav system compared to the unaided INS in a standard rental van, following trajectory illustrated in Fig.~\ref{fig:trajectory_maps_all}c. Over the first few kilometers the system undergoes online calibration, learning about the magnetic characteristics of the vehicle. Once a level of confidence has been reached, the system begins to provide navigation corrections.}
\label{fig:position_error_ground_trial}
\end{figure}

Returning to the ground, Fig.~\ref{fig:position_error_ground_trial} shows the results of a ground-based trial, for a loop trajectory overlaid with the magnetic anomaly map. Overall, the results are qualitatively similar to flight trials, showing smaller but consistent relative advantages due to the extreme noise experienced by the magnetometers. The measured magnetic anomalies at ground level are approximately $600$\,nT full range over the trip, and the measured magnetic noise inside the van reached up to $30,000$\,nT. Despite the unfavorably high $\sim$50$\times$ noise-to-signal ratio, the data showcase how MagNav is able to deliver a $\sim7 \times$ improvement in positioning accuracy at the end of a $17.5$\,km path, with MagNav positioning accuracy reduced to just $180$\,m. To the best of our knowledge there has not previously been a successful ground-based trial of MagNav using independently compiled magnetic maps.

\section*{Discussion}
In these field trials, across multiple airborne and ground-based missions, we validate that quantum-assured MagNav can deliver more accurate and precise positioning than a strategic-grade INS under real-world operating conditions relevant to defense and civilian applications.  We compare against a high-end strategic-grade INS with comparable size, weight, and power to the MagNav system and outperform by up to $\sim$46$\times$, with typical results showcasing $11-38\times$ outperformance. We reiterate that this choice of classical comparator is most relevant because of the similar operating conditions supported by both the INS and MagNav: all-weather, passive, and unjammable/unspoofable operation. 

These results go far beyond previous public-domain demonstrations of MagNav of which we are aware; we acknowledge that clandestine demonstrations may exist of which we do not have knowledge. First, these results include the best reported positioning accuracy with MagNav; our best-accuracy trial achieves accuracy just $0.006\%$ of the distance traveled, exceeding most previously-reported alternative-navigation solutions. In our MagNav trials this positioning accuracy was achieved without any requirements for clear line-of-sight to ground/sky or active emission of any signal. Next, we believe that these results constitute the only published demonstration of MagNav directly outperforming a strategic-grade INS for the same task using any form of quasi-real-time correction. In addition, these flight trials represent the first public-domain demonstration of MagNav without the requirement of specialized in-flight training maneuvers.

Further, to the best of our knowledge our successful ground-based trials themselves represent a world-first demonstration; the trials we conducted rely exclusively on independently-generated and publicly-available magnetic anomaly maps produced under different conditions than the trial, and we also did not make any assumptions about constraints on the vehicle's path. The only comparable prior field test of which we are aware was significantly more narrow: Shockley and Raquet \cite{shockley2014} used a vector magnetometer on a ground vehicle to provide a magnetic-positioning solution. However, they magnetically pre-mapped the route rather than relying on a public map, meaning their map included large anthropogenic magnetic features such as power lines and did not rely on the magnetic anomaly field alone. Additionally, their filter made the assumption that the vehicle was constrained to remain on the road system.

Our detailed investigations have revealed several key insights relevant to deployment of quantum-assured MagNav in real operational environments. First, we have validated in-flight performance of our magnetometers, achieving superior performance to alternative solutions in flight tests experiencing temperatures in the range $[12, 40]^{\circ}$C. Overall resilience testing showed strong results for our quantum magnetometers, contrasted with repeated failures and data-stream instabilities observed in readily-available COTS alternatives. Flight tests were also conducted using steep climb and bank maneuvers that dramatically exceed those acceptable in commercial aircraft, validating performance under a range of flight envelope conditions, and confirming laboratory vibration testing up to $\sim6g$. 

Next, we have performed direct comparisons between the cold- and warm-start denoising modes in flight trials, demonstrating comparable performance over flights of several hundred kilometers (Fig.~\ref{fig:cold_vs_warm}). This finding has direct operational relevance as it enables full re-optimization of the MagNav navigation algorithm under dramatically-changed conditions with no special training datasets, calibration, or vehicle maneuvers required. The fact that warm-start also works using pretrained coefficients opens the possibility to have immediate access to effective MagNav with no online learning in circumstances where a vehicle may face immediate GNSS denial or jamming at the start of a mission: e.g. flight take-off from an airport in a region known for GNSS denial.

Additional simulation-based tests of our MagNav software stack also substantiate the robustness of our approach under realistic environmental conditions. The cold-start filter can not only continuously adapt to changing payloads or flight conditions, but it is also resilient against changing latitudes. This is crucial as a potential MagNav failure mode arises from the fact that a large change in latitude requires a full recalibration of Tolles-Lawson coefficients or retraining of data-driven AI algorithms to account for the different value of Earth's field~\cite{noriega2011, canciani-f16,feng2022, liu2022}. Our use of an iterative, adaptive, physics-based algorithm instead imparts resilience against large constant offsets in the Earth field. This enables full operation in flight paths with significant meridional projections, and also offsets concerns about slow drifts in the location of Earth's magnetic poles which have otherwise interfered with magnetic heading corrections in ILS systems.

Finally, our tests have validated that our algorithms are powerful enough to enable successful denoising and map matching at operationally-relevant altitudes with the quantum magnetometers positioned inside the cargo bay of the aircraft, mounted adjacent to electronics. This enables ``drop-in'' solutions to immediately deliver capability. However, the observation that the best trial result (in terms of improvement over the INS) was achieved using externally-mounted quantum magnetometers presents an opportunity to explore enhanced MagNav performance by engineering magnetometers into vehicle design. For instance, the very modest $\sim63$~cc volume of Q-CTRL's current generation of quantum magnetometers enables mounting on wingtips or in a tailcone in order to present a lower-noise environment for augmented operations at high altitudes.

The optimal sensor architecture for MagNav remains a work in progress. We chose to perform our demonstrations using one quantum scalar magnetometer and one classical vector magnetometer. This matches standard practice in geophysical survey and magnetic navigation~\cite{canciani2,canciani2016}, where quantum optically-pumped magnetometers are selected because of their improved sensitivity and stability. However, it remains an open question whether multiple quantum scalar magnetometers and/or classical magnetometers could be leveraged to achieve an improved position accuracy. 

Magnetic navigation has only a few limitations: magnetic maps are not available everywhere; position-fix updates can become less precise in regions where there is little to no magnetic variation; and there are rare events when extraordinarily powerful solar events can mask magnetic features. Fortunately, all of these limitations can be appropriately identified and accounted for in the map-matching algorithm, ensuring the position fix remains accurate, while uncertainty may increase.

We also believe it is relevant to explore how MagNav can be integrated with complementary alternative navigation solutions to provide robust, high-performance alternatives to GNSS. The limitations of MagNav, articulated above, are complementary to those of optical position-fixing solutions. Through fusion of position estimates emerging from both technologies in parallel, it may be possible to more accurately determine positioning information and ensure coverage of a wider range of conditions. We have focused here on the operation of a standalone MagNav solution in order to ensure robust navigational solutions in circumstances where optical position-fixing is not viable.

Considerable work remains to be done to validate this technology at commercial airline cruising altitudes, under highly-dynamic maneuvers as encountered in military aircraft, and over water where magnetic anomalies are often smaller than over land. In particular, the availability and quality of magnetic anomaly map coverage in critical operational areas remains to be validated or augmented as needed. Public-domain global maps have limited resolution (2 arc minutes, or approximately 3\,km), are given at reference heights that limit usability (e.g. 6\,km), and have limited ocean coverage. Furthermore, as these are harmonized collections of individual surveys, some of which were performed before the advent of GNSS, their errors can be large. Ongoing map development and expansion of map coverage will be an important part of ensuring broad applicability of this technology.

\appendix
\section*{Appendix}
Additional performance information on the strategic-grade INS used for MagNav comparisons.

\begin{table}[h]
    \centering
    \begin{tabular}{l l l}
       Sensor & Accelerometers & Gyroscopes \\
       \hline
        Range & $\pm$ 15\,g & $\pm$ 490$^{\circ}$/\,s \\
        Bias nstability & 7\,$\mu$g & 0.001$^{\circ}$/\,hr \\ 
        Initial bias & $<$ 100\,$\mu$g & $<$ 0.01$^{\circ}$/\,hr \\
        Initial scale error & 340 ppm & 80 ppm \\
        Scale factor stability & 100 ppm & 10 ppm  \\
        Non-linearity & 150 ppm  & 10 ppm  \\
        Cross-axis alignment error & $<$ 0.001$^{\circ}$  & $<$ 0.001$^{\circ}$  \\
        Noise density & 30\,$\mu$g/$\sqrt{\text{Hz}}$  & 0.06$^{\circ}$/\,hr/$\sqrt{\text{Hz}}$  \\
        Random walk (VRW/ARW) & 17 mm/s/$\sqrt{\text{hr}}$  &  0.001$^{\circ}$/$\sqrt{\text{hr}}$  \\
        Bandwidth & 300\,Hz  & 400\,Hz  \\
    \end{tabular}
    \caption{IMU sensor specifications of the INS used for GNSS-free navigation in these trials.}
    \label{tab:INSspecs}
\end{table}

\bibliography{magnav_bibliography}

\begin{thebibliography}{33}%
\makeatletter
\providecommand \@ifxundefined [1]{%
 \@ifx{#1\undefined}
}%
\providecommand \@ifnum [1]{%
 \ifnum #1\expandafter \@firstoftwo
 \else \expandafter \@secondoftwo
 \fi
}%
\providecommand \@ifx [1]{%
 \ifx #1\expandafter \@firstoftwo
 \else \expandafter \@secondoftwo
 \fi
}%
\providecommand \natexlab [1]{#1}%
\providecommand \enquote  [1]{``#1''}%
\providecommand \bibnamefont  [1]{#1}%
\providecommand \bibfnamefont [1]{#1}%
\providecommand \citenamefont [1]{#1}%
\providecommand \href@noop [0]{\@secondoftwo}%
\providecommand \href [0]{\begingroup \@sanitize@url \@href}%
\providecommand \@href[1]{\@@startlink{#1}\@@href}%
\providecommand \@@href[1]{\endgroup#1\@@endlink}%
\providecommand \@sanitize@url [0]{\catcode `\\12\catcode `\$12\catcode `\&12\catcode `\#12\catcode `\^12\catcode `\_12\catcode `\%12\relax}%
\providecommand \@@startlink[1]{}%
\providecommand \@@endlink[0]{}%
\providecommand \url  [0]{\begingroup\@sanitize@url \@url }%
\providecommand \@url [1]{\endgroup\@href {#1}{\urlprefix }}%
\providecommand \urlprefix  [0]{URL }%
\providecommand \Eprint [0]{\href }%
\providecommand \doibase [0]{http://dx.doi.org/}%
\providecommand \selectlanguage [0]{\@gobble}%
\providecommand \bibinfo  [0]{\@secondoftwo}%
\providecommand \bibfield  [0]{\@secondoftwo}%
\providecommand \translation [1]{[#1]}%
\providecommand \BibitemOpen [0]{}%
\providecommand \bibitemStop [0]{}%
\providecommand \bibitemNoStop [0]{.\EOS\space}%
\providecommand \EOS [0]{\spacefactor3000\relax}%
\providecommand \BibitemShut  [1]{\csname bibitem#1\endcsname}%
\let\auto@bib@innerbib\@empty
\bibitem [{\citenamefont {Hofmann-Wellenhof}\ \emph {et~al.}(2008)\citenamefont {Hofmann-Wellenhof}, \citenamefont {Lichtenegger},\ and\ \citenamefont {Wasle}}]{Bernhard2008}%
  \BibitemOpen
  \bibfield  {author} {\bibinfo {author} {\bibfnamefont {B.}~\bibnamefont {Hofmann-Wellenhof}}, \bibinfo {author} {\bibfnamefont {H.}~\bibnamefont {Lichtenegger}}, \ and\ \bibinfo {author} {\bibfnamefont {E.}~\bibnamefont {Wasle}},\ }\href@noop {} {\emph {\bibinfo {title} {{GNSS - Global Navigation Satellite Systems : GPS, GLONASS, Galileo \& more (Japanese language)}}}},\ \bibinfo {edition} {{Translation of the first edition of the Springer Publishing Com}}\ ed.\ (\bibinfo  {publisher} {Springer Verlag},\ \bibinfo {address} {Deutschland},\ \bibinfo {year} {2008})\BibitemShut {NoStop}%
\bibitem [{\citenamefont {Montenbruck}\ \emph {et~al.}(2017)\citenamefont {Montenbruck}, \citenamefont {Steigenberger}, \citenamefont {Prange}, \citenamefont {Deng}, \citenamefont {Zhao}, \citenamefont {Perosanz}, \citenamefont {Romero}, \citenamefont {Noll}, \citenamefont {St\"urze}, \citenamefont {Weber}, \citenamefont {Schmid}, \citenamefont {MacLeod},\ and\ \citenamefont {Schaer}}]{Montenbruck2017}%
  \BibitemOpen
  \bibfield  {author} {\bibinfo {author} {\bibfnamefont {O.}~\bibnamefont {Montenbruck}}, \bibinfo {author} {\bibfnamefont {P.}~\bibnamefont {Steigenberger}}, \bibinfo {author} {\bibfnamefont {L.}~\bibnamefont {Prange}}, \bibinfo {author} {\bibfnamefont {Z.}~\bibnamefont {Deng}}, \bibinfo {author} {\bibfnamefont {Q.}~\bibnamefont {Zhao}}, \bibinfo {author} {\bibfnamefont {F.}~\bibnamefont {Perosanz}}, \bibinfo {author} {\bibfnamefont {I.}~\bibnamefont {Romero}}, \bibinfo {author} {\bibfnamefont {C.}~\bibnamefont {Noll}}, \bibinfo {author} {\bibfnamefont {A.}~\bibnamefont {St\"urze}}, \bibinfo {author} {\bibfnamefont {G.}~\bibnamefont {Weber}}, \bibinfo {author} {\bibfnamefont {R.}~\bibnamefont {Schmid}}, \bibinfo {author} {\bibfnamefont {K.}~\bibnamefont {MacLeod}}, \ and\ \bibinfo {author} {\bibfnamefont {S.}~\bibnamefont {Schaer}},\ }\bibfield  {title} {\enquote {\bibinfo {title} {The {Multi-GNSS} experiment ({MGEX}) of the international {GNSS} service ({IGS}) – achievements, prospects and challenges},}\
  }\href {\doibase https://doi.org/10.1016/j.asr.2017.01.011} {\bibfield  {journal} {\bibinfo  {journal} {Advances in Space Research}\ }\textbf {\bibinfo {volume} {59}},\ \bibinfo {pages} {1671--1697} (\bibinfo {year} {2017})}\BibitemShut {NoStop}%
\bibitem [{\citenamefont {of~Engineering (Great~Britain)}(2011)}]{royal2011global}%
  \BibitemOpen
  \bibfield  {author} {\bibinfo {author} {\bibfnamefont {Royal~Academy}\ \bibnamefont {of~Engineering (Great~Britain)}},\ }\href {https://books.google.com.au/books?id=PJSSZwEACAAJ} {\emph {\bibinfo {title} {Global Navigation Space Systems: Reliance and Vulnerabilities}}}\ (\bibinfo  {publisher} {Royal Academy of Engineering},\ \bibinfo {year} {2011})\BibitemShut {NoStop}%
\bibitem [{\citenamefont {Titterton}\ and\ \citenamefont {Weston}(2004)}]{titterton}%
  \BibitemOpen
  \bibfield  {author} {\bibinfo {author} {\bibfnamefont {D.~H.}\ \bibnamefont {Titterton}}\ and\ \bibinfo {author} {\bibfnamefont {J.~L.}\ \bibnamefont {Weston}},\ }\href@noop {} {\emph {\bibinfo {title} {Strapdown Inertial Navigation Technology, 2nd Edition}}}\ (\bibinfo  {publisher} {The Institution of Electrical Engineers},\ \bibinfo {year} {2004})\BibitemShut {NoStop}%
\bibitem [{\citenamefont {Son}\ \emph {et~al.}(2018)\citenamefont {Son}, \citenamefont {Rhee},\ and\ \citenamefont {Seo}}]{Son:2018}%
  \BibitemOpen
  \bibfield  {author} {\bibinfo {author} {\bibfnamefont {P.-W.}\ \bibnamefont {Son}}, \bibinfo {author} {\bibfnamefont {J.~H.}\ \bibnamefont {Rhee}}, \ and\ \bibinfo {author} {\bibfnamefont {J.}~\bibnamefont {Seo}},\ }\bibfield  {title} {\enquote {\bibinfo {title} {Novel multichain-based {L}oran positioning algorithm for resilient navigation},}\ }\href {\doibase 10.1109/TAES.2017.2762438} {\bibfield  {journal} {\bibinfo  {journal} {IEEE Transactions on Aerospace and Electronic Systems}\ }\textbf {\bibinfo {volume} {54}},\ \bibinfo {pages} {666--679} (\bibinfo {year} {2018})}\BibitemShut {NoStop}%
\bibitem [{\citenamefont {Zafari}\ \emph {et~al.}(2019)\citenamefont {Zafari}, \citenamefont {Gkelias},\ and\ \citenamefont {Leung}}]{zafari2019surveyindoorlocalizationsystems}%
  \BibitemOpen
  \bibfield  {author} {\bibinfo {author} {\bibfnamefont {F.}~\bibnamefont {Zafari}}, \bibinfo {author} {\bibfnamefont {A.}~\bibnamefont {Gkelias}}, \ and\ \bibinfo {author} {\bibfnamefont {K.}~\bibnamefont {Leung}},\ }\href {https://arxiv.org/abs/1709.01015} {\enquote {\bibinfo {title} {A survey of indoor localization systems and technologies},}\ } (\bibinfo {year} {2019}),\ \Eprint {http://arxiv.org/abs/1709.01015} {arXiv:1709.01015 [cs.NI]} \BibitemShut {NoStop}%
\bibitem [{\citenamefont {Dawson}\ \emph {et~al.}(2022)\citenamefont {Dawson}, \citenamefont {Rashed}, \citenamefont {Abdelfatah},\ and\ \citenamefont {Noureldin}}]{RadarDawson2022}%
  \BibitemOpen
  \bibfield  {author} {\bibinfo {author} {\bibfnamefont {E.}~\bibnamefont {Dawson}}, \bibinfo {author} {\bibfnamefont {M.~A.}\ \bibnamefont {Rashed}}, \bibinfo {author} {\bibfnamefont {W.}~\bibnamefont {Abdelfatah}}, \ and\ \bibinfo {author} {\bibfnamefont {A.}~\bibnamefont {Noureldin}},\ }\bibfield  {title} {\enquote {\bibinfo {title} {{Radar-Based Multisensor Fusion for Uninterrupted Reliable Positioning in GNSS-Denied Environments}},}\ }\href {\doibase 10.1109/TITS.2022.3202139} {\bibfield  {journal} {\bibinfo  {journal} {IEEE Transactions on Intelligent Transportation Systems}\ }\textbf {\bibinfo {volume} {23}},\ \bibinfo {pages} {23384--23398} (\bibinfo {year} {2022})}\BibitemShut {NoStop}%
\bibitem [{\citenamefont {Zhang}\ and\ \citenamefont {Singh}(2017)}]{LIDARZhang2017}%
  \BibitemOpen
  \bibfield  {author} {\bibinfo {author} {\bibfnamefont {J.}~\bibnamefont {Zhang}}\ and\ \bibinfo {author} {\bibfnamefont {S.}~\bibnamefont {Singh}},\ }\bibfield  {title} {\enquote {\bibinfo {title} {Low-drift and real-time lidar odometry and mapping},}\ }\href {\doibase 10.1007/s10514-016-9548-2} {\bibfield  {journal} {\bibinfo  {journal} {Auton. Robots}\ }\textbf {\bibinfo {volume} {41}},\ \bibinfo {pages} {401–416} (\bibinfo {year} {2017})}\BibitemShut {NoStop}%
\bibitem [{\citenamefont {Bijjahalli}\ \emph {et~al.}(2020)\citenamefont {Bijjahalli}, \citenamefont {Sabatini},\ and\ \citenamefont {Gardi}}]{Bijjahalli:2020}%
  \BibitemOpen
  \bibfield  {author} {\bibinfo {author} {\bibfnamefont {S.}~\bibnamefont {Bijjahalli}}, \bibinfo {author} {\bibfnamefont {R.}~\bibnamefont {Sabatini}}, \ and\ \bibinfo {author} {\bibfnamefont {A.}~\bibnamefont {Gardi}},\ }\bibfield  {title} {\enquote {\bibinfo {title} {{Advances in intelligent and autonomous navigation systems for small UAS}},}\ }\href {\doibase https://doi.org/10.1016/j.paerosci.2020.100617} {\bibfield  {journal} {\bibinfo  {journal} {Progress in Aerospace Sciences}\ }\textbf {\bibinfo {volume} {115}},\ \bibinfo {pages} {100617} (\bibinfo {year} {2020})}\BibitemShut {NoStop}%
\bibitem [{\citenamefont {Groves}(2013)}]{groves2013}%
  \BibitemOpen
  \bibfield  {author} {\bibinfo {author} {\bibfnamefont {P.~D.}\ \bibnamefont {Groves}},\ }\href@noop {} {\emph {\bibinfo {title} {Principles of GNSS, Inertial, and Multisensor Integrated Navigation Systems}}}\ (\bibinfo  {publisher} {Artech House},\ \bibinfo {year} {2013})\BibitemShut {NoStop}%
\bibitem [{\citenamefont {Balamurugan}\ \emph {et~al.}(2016)\citenamefont {Balamurugan}, \citenamefont {Valarmathi},\ and\ \citenamefont {Naidu}}]{Balamurugan2016SurveyOU}%
  \BibitemOpen
  \bibfield  {author} {\bibinfo {author} {\bibfnamefont {G.}~\bibnamefont {Balamurugan}}, \bibinfo {author} {\bibfnamefont {J.}~\bibnamefont {Valarmathi}}, \ and\ \bibinfo {author} {\bibfnamefont {V.~P.~S.}\ \bibnamefont {Naidu}},\ }\bibfield  {title} {\enquote {\bibinfo {title} {{Survey on UAV navigation in GPS denied environments}},}\ }\href {https://api.semanticscholar.org/CorpusID:34905197} {\bibfield  {journal} {\bibinfo  {journal} {{2016 International Conference on Signal Processing, Communication, Power and Embedded System (SCOPES)}}\ ,\ \bibinfo {pages} {198--204}} (\bibinfo {year} {2016})}\BibitemShut {NoStop}%
\bibitem [{\citenamefont {Canciani}\ and\ \citenamefont {Raquet}(2020)}]{canciani2016}%
  \BibitemOpen
  \bibfield  {author} {\bibinfo {author} {\bibfnamefont {A.~J.}\ \bibnamefont {Canciani}}\ and\ \bibinfo {author} {\bibfnamefont {J.}~\bibnamefont {Raquet}},\ }\bibfield  {title} {\enquote {\bibinfo {title} {Absolute positioning using the earth’s magnetic anomaly field},}\ }\href@noop {} {\bibfield  {journal} {\bibinfo  {journal} {Journal of The Institute of Navigation}\ }\textbf {\bibinfo {volume} {63}},\ \bibinfo {pages} {111} (\bibinfo {year} {2020})}\BibitemShut {NoStop}%
\bibitem [{\citenamefont {P.~Alken}(2022)}]{IGRF}%
  \BibitemOpen
  \bibfield  {author} {\bibinfo {author} {\bibfnamefont {C.~D. Beggan~\emph{et al}.}\ \bibnamefont {P.~Alken}, \bibfnamefont {E.~Thébault}},\ }\bibfield  {title} {\enquote {\bibinfo {title} {International geomagnetic reference field: the 13th generation},}\ }\href@noop {} {\bibfield  {journal} {\bibinfo  {journal} {Earth, Planets and Space}\ }\textbf {\bibinfo {volume} {73}},\ \bibinfo {pages} {11} (\bibinfo {year} {2022})}\BibitemShut {NoStop}%
\bibitem [{\citenamefont {B.~Meyer}(2017)}]{EMAG2v3}%
  \BibitemOpen
  \bibfield  {author} {\bibinfo {author} {\bibfnamefont {R.~Saltus}\ \bibnamefont {B.~Meyer}, \bibfnamefont {A.~Chulliat}},\ }\bibfield  {title} {\enquote {\bibinfo {title} {Derivation and error analysis of the earth magnetic anomaly grid at 2 arc min resolution version 3},}\ }\href@noop {} {\bibfield  {journal} {\bibinfo  {journal} {Geochemistry, Geophysics and Geosystems}\ }\textbf {\bibinfo {volume} {12}},\ \bibinfo {pages} {4522} (\bibinfo {year} {2017})}\BibitemShut {NoStop}%
\bibitem [{\citenamefont {Choi}\ \emph {et~al.}()\citenamefont {Choi}, \citenamefont {Dyment}, \citenamefont {Lesur}, \citenamefont {Reyes}, \citenamefont {Catalan}, \citenamefont {Ishihara}, \citenamefont {Litvinova},\ and\ \citenamefont {Hamoudi}}]{WDMAM}%
  \BibitemOpen
  \bibfield  {author} {\bibinfo {author} {\bibfnamefont {Y.}~\bibnamefont {Choi}}, \bibinfo {author} {\bibfnamefont {J.}~\bibnamefont {Dyment}}, \bibinfo {author} {\bibfnamefont {V.}~\bibnamefont {Lesur}}, \bibinfo {author} {\bibfnamefont {G.}~\bibnamefont {Reyes}}, \bibinfo {author} {\bibfnamefont {M.}~\bibnamefont {Catalan}}, \bibinfo {author} {\bibfnamefont {T.}~\bibnamefont {Ishihara}}, \bibinfo {author} {\bibfnamefont {T.}~\bibnamefont {Litvinova}}, \ and\ \bibinfo {author} {\bibfnamefont {M.}~\bibnamefont {Hamoudi}},\ }\href@noop {} {\enquote {\bibinfo {title} {{World Digital Magnetic Anomaly Map version 2.2}},}\ }\bibinfo {note} {{www.wdmam.org}}\BibitemShut {NoStop}%
\bibitem [{\citenamefont {Langel}\ and\ \citenamefont {Hinze}(1998)}]{langel1998}%
  \BibitemOpen
  \bibfield  {author} {\bibinfo {author} {\bibfnamefont {R.~A.}\ \bibnamefont {Langel}}\ and\ \bibinfo {author} {\bibfnamefont {W.~J.}\ \bibnamefont {Hinze}},\ }\href@noop {} {\emph {\bibinfo {title} {The Magnetic Field of the Earth’s Lithosphere: The Satellite Perspective}}}\ (\bibinfo  {publisher} {Cambridge Univ. Press},\ \bibinfo {year} {1998})\BibitemShut {NoStop}%
\bibitem [{\citenamefont {McNeil}(2022)}]{mcneil2022}%
  \BibitemOpen
  \bibfield  {author} {\bibinfo {author} {\bibfnamefont {A.~J.}\ \bibnamefont {McNeil}},\ }\emph {\bibinfo {title} {Magnetic Anomaly Absolute Positioning For Hypersonic Aircraft}},\ \href@noop {} {Master's thesis},\ \bibinfo  {school} {Air Force Institute of Technology} (\bibinfo {year} {2022})\BibitemShut {NoStop}%
\bibitem [{\citenamefont {\emph{et al}.}(2023)}]{greentree2023}%
  \BibitemOpen
  \bibfield  {author} {\bibinfo {author} {\bibfnamefont {X.~Wang}\ \bibnamefont {\emph{et al}.}},\ }\href@noop {} {\enquote {\bibinfo {title} {Quantum diamond magnetometry for navigation in {GNSS} denied environments},}\ } (\bibinfo {year} {2023}),\ \Eprint {http://arxiv.org/abs/2302.06187} {arXiv:2302.06187} \BibitemShut {NoStop}%
\bibitem [{\citenamefont {Gupta}(2024)}]{gupta}%
  \BibitemOpen
  \bibfield  {author} {\bibinfo {author} {\bibfnamefont {A.}~\bibnamefont {Gupta}},\ }\bibfield  {title} {\enquote {\bibinfo {title} {Lower bounds on magnetic navigation performance as a function of magnetic anomaly map quality},}\ }in\ \href@noop {} {\emph {\bibinfo {booktitle} {AIAA DATC/IEEE 43rd Digital Avionics Systems Conference}}}\ (\bibinfo {year} {2024})\BibitemShut {NoStop}%
\bibitem [{\citenamefont {Tkhorenko}\ and\ \citenamefont {Karshakov}(2022)}]{tkhorenko}%
  \BibitemOpen
  \bibfield  {author} {\bibinfo {author} {\bibfnamefont {M.}~\bibnamefont {Tkhorenko}}\ and\ \bibinfo {author} {\bibfnamefont {E.}~\bibnamefont {Karshakov}},\ }\bibfield  {title} {\enquote {\bibinfo {title} {Estimating the potential accuracy of magnetic navigation based on magnetic survey data},}\ }in\ \href@noop {} {\emph {\bibinfo {booktitle} {29th Saint Petersburg International Conference on Integrated Navigation Systems}}}\ (\bibinfo {year} {2022})\BibitemShut {NoStop}%
\bibitem [{\citenamefont {\emph{et al}.}(2006)}]{wilson2006}%
  \BibitemOpen
  \bibfield  {author} {\bibinfo {author} {\bibfnamefont {J.M.~Wilson}\ \bibnamefont {\emph{et al}.}},\ }\bibfield  {title} {\enquote {\bibinfo {title} {Passive navigation using local magnetic field variations},}\ }in\ \href@noop {} {\emph {\bibinfo {booktitle} {Proceedings of the 2006 National Technical Meeting of The Institute of Navigation, Monterey, CA}}}\ (\bibinfo {year} {2006})\ pp.\ \bibinfo {pages} {770--779}\BibitemShut {NoStop}%
\bibitem [{\citenamefont {Canciani}\ and\ \citenamefont {Raquet}(2016)}]{cancianiMapQuality}%
  \BibitemOpen
  \bibfield  {author} {\bibinfo {author} {\bibfnamefont {A.~J.}\ \bibnamefont {Canciani}}\ and\ \bibinfo {author} {\bibfnamefont {J.~F.}\ \bibnamefont {Raquet}},\ }\bibfield  {title} {\enquote {\bibinfo {title} {Magnetic anomaly navigation accuracy with respect to map quality and altitude},}\ }in\ \href@noop {} {\emph {\bibinfo {booktitle} {Proceedings of the 2016 International Technical Meeting of The Institute of Navigation}}}\ (\bibinfo {year} {2016})\ p.\ \bibinfo {pages} {1241}\BibitemShut {NoStop}%
\bibitem [{\citenamefont {Canciani}\ and\ \citenamefont {Raquet}(2017)}]{canciani2}%
  \BibitemOpen
  \bibfield  {author} {\bibinfo {author} {\bibfnamefont {A.~J.}\ \bibnamefont {Canciani}}\ and\ \bibinfo {author} {\bibfnamefont {J.}~\bibnamefont {Raquet}},\ }\bibfield  {title} {\enquote {\bibinfo {title} {Airborne magnetic anomaly navigation},}\ }\href@noop {} {\bibfield  {journal} {\bibinfo  {journal} {IEEE Transactions on Aerospace and Electronic Systems}\ }\textbf {\bibinfo {volume} {53}},\ \bibinfo {pages} {67} (\bibinfo {year} {2017})}\BibitemShut {NoStop}%
\bibitem [{\citenamefont {Canciani}(2022)}]{canciani-f16}%
  \BibitemOpen
  \bibfield  {author} {\bibinfo {author} {\bibfnamefont {A.~J.}\ \bibnamefont {Canciani}},\ }\bibfield  {title} {\enquote {\bibinfo {title} {Magnetic navigation on an {F}-16 aircraft using online calibration},}\ }\href@noop {} {\bibfield  {journal} {\bibinfo  {journal} {IEEE Transactions on Aerospace and Electronic Systems}\ }\textbf {\bibinfo {volume} {58}},\ \bibinfo {pages} {420} (\bibinfo {year} {2022})}\BibitemShut {NoStop}%
\bibitem [{\citenamefont {Lee}\ and\ \citenamefont {Canciani}(2020)}]{leecanciani}%
  \BibitemOpen
  \bibfield  {author} {\bibinfo {author} {\bibfnamefont {T.~N.}\ \bibnamefont {Lee}}\ and\ \bibinfo {author} {\bibfnamefont {A.~J.}\ \bibnamefont {Canciani}},\ }\bibfield  {title} {\enquote {\bibinfo {title} {{MagSLAM: Aerial simultaneous localization and mapping using Earth's magnetic anomaly field}},}\ }\href@noop {} {\bibfield  {journal} {\bibinfo  {journal} {Navigation}\ }\textbf {\bibinfo {volume} {67}},\ \bibinfo {pages} {95} (\bibinfo {year} {2020})}\BibitemShut {NoStop}%
\bibitem [{\citenamefont {Budker}\ and\ \citenamefont {Kimble}(2013)}]{Budker:2013}%
  \BibitemOpen
  \bibfield  {author} {\bibinfo {author} {\bibfnamefont {Dmitry}\ \bibnamefont {Budker}}\ and\ \bibinfo {author} {\bibfnamefont {Derek F.~Jackson}\ \bibnamefont {Kimble}},\ }\href@noop {} {\emph {\bibinfo {title} {Optical Magnetometry}}}\ (\bibinfo  {publisher} {Cambridge University Press},\ \bibinfo {year} {2013})\BibitemShut {NoStop}%
\bibitem [{\citenamefont {Tolles}\ and\ \citenamefont {Lawson}(1950)}]{tolles1950}%
  \BibitemOpen
  \bibfield  {author} {\bibinfo {author} {\bibfnamefont {W.~E.}\ \bibnamefont {Tolles}}\ and\ \bibinfo {author} {\bibfnamefont {J.~D.}\ \bibnamefont {Lawson}},\ }\href@noop {} {\enquote {\bibinfo {title} {Magnetic compensation of {MAD} equipped aircraft},}\ }\bibinfo {howpublished} {Report 201-1} (\bibinfo {year} {1950}),\ \bibinfo {note} {{A}irborne Instruments Laboratory Inc}\BibitemShut {NoStop}%
\bibitem [{\citenamefont {Gnadt}\ \emph {et~al.}(2022)\citenamefont {Gnadt}, \citenamefont {Wollaber},\ and\ \citenamefont {Nielsen}}]{gnadt2022}%
  \BibitemOpen
  \bibfield  {author} {\bibinfo {author} {\bibfnamefont {Albert~R.}\ \bibnamefont {Gnadt}}, \bibinfo {author} {\bibfnamefont {Allan~B.}\ \bibnamefont {Wollaber}}, \ and\ \bibinfo {author} {\bibfnamefont {Aaron~P.}\ \bibnamefont {Nielsen}},\ }\href {https://arxiv.org/abs/2212.09899} {\enquote {\bibinfo {title} {Derivation and extensions of the tolles-lawson model for aeromagnetic compensation},}\ } (\bibinfo {year} {2022}),\ \Eprint {http://arxiv.org/abs/2212.09899} {arXiv:2212.09899 [physics.app-ph]} \BibitemShut {NoStop}%
\bibitem [{Total Magnetic Intensity (TMI) Grid of Australia (2019), 7th Edition()}]{TMIAustralia}%
  \BibitemOpen
  Total Magnetic Intensity (TMI) Grid of Australia (2019), 7th Edition,\ \href {ecat.ga.gov.au/geonetwork/srv/api/records/7c38a2b2-28e4-4c79-b0ce-517d9861e20d} {} (\bibinfo {year} {2019})\BibitemShut {NoStop}%
\bibitem [{\citenamefont {Shockley}\ and\ \citenamefont {Raquet}(2014)}]{shockley2014}%
  \BibitemOpen
  \bibfield  {author} {\bibinfo {author} {\bibfnamefont {J.~A.}\ \bibnamefont {Shockley}}\ and\ \bibinfo {author} {\bibfnamefont {J.~F.}\ \bibnamefont {Raquet}},\ }\bibfield  {title} {\enquote {\bibinfo {title} {Navigation of ground vehicles using magnetic field variations},}\ }\href@noop {} {\bibfield  {journal} {\bibinfo  {journal} {Journal of The Institute of Navigation}\ }\textbf {\bibinfo {volume} {61}},\ \bibinfo {pages} {4} (\bibinfo {year} {2014})}\BibitemShut {NoStop}%
\bibitem [{\citenamefont {Noriega-Besga}(2011)}]{noriega2011}%
  \BibitemOpen
  \bibfield  {author} {\bibinfo {author} {\bibfnamefont {G.}~\bibnamefont {Noriega-Besga}},\ }\bibfield  {title} {\enquote {\bibinfo {title} {Performance measures in aeromagnetic compensation},}\ }\href@noop {} {\bibfield  {journal} {\bibinfo  {journal} {The Leading Edge}\ }\textbf {\bibinfo {volume} {30}},\ \bibinfo {pages} {1122} (\bibinfo {year} {2011})}\BibitemShut {NoStop}%
\bibitem [{\citenamefont {\emph{et al.}}(2022)}]{feng2022}%
  \BibitemOpen
  \bibfield  {author} {\bibinfo {author} {\bibfnamefont {Y.~Feng}\ \bibnamefont {\emph{et al.}}},\ }\bibfield  {title} {\enquote {\bibinfo {title} {An improved aeromagnetic compensation method robust to geomagnetic gradient},}\ }\href@noop {} {\bibfield  {journal} {\bibinfo  {journal} {Appl. Sci.}\ }\textbf {\bibinfo {volume} {12}},\ \bibinfo {pages} {1490} (\bibinfo {year} {2022})}\BibitemShut {NoStop}%
\bibitem [{\citenamefont {\emph{et al}.}(2022)}]{liu2022}%
  \BibitemOpen
  \bibfield  {author} {\bibinfo {author} {\bibfnamefont {X.~Liu}\ \bibnamefont {\emph{et al}.}},\ }\bibfield  {title} {\enquote {\bibinfo {title} {Application of an improved calibration flight scheme in aeromagnetic interference compensation},}\ }\href@noop {} {\bibfield  {journal} {\bibinfo  {journal} {JGR Solid Earth}\ }\textbf {\bibinfo {volume} {127}} (\bibinfo {year} {2022})}\BibitemShut {NoStop}%
\end{thebibliography}%

\end{document}